\documentclass [12pt]{article}
\usepackage {graphicx}
\usepackage{psfrag}
\usepackage{vmargin}  
\setmarginsrb{2.5cm}{2.5cm}{2.5cm}{2.5cm}{0cm}{0.cm}{0cm}{1cm}

\begin{document}

\begin{flushright}
PCCF--RI--0205\\
7.08.2002
\end{flushright}
\vfil

\begin{center}
{\Large \bf LIMITS ON THE MIXING OF TAU NEUTRINO\\
TO HEAVY NEUTRINOS \par}
\vskip 3em
{
\lineskip .75em
\begin{tabular}[t]{c}
J. Orloff$^{1)}$, A. Rozanov$^{2)}$ and C. Santoni$^{1)}$\\
\\
$^{1)}$ Laboratoire de Physique corpusculaire de Clermont-Ferrand 
IN2P3/CNRS,\\
	Universit\'{e} Blaise Pascal, Clermont-Ferrand, France\\
\\
$^{2)}$ Centre de Physique des Particules de Marseille IN2P3/CNRS,\\
	Universit\'{e} M\'{e}diterran\'{e}e, Marseille, France
\end{tabular}
}
\end{center}

\vfil

\abstract{Limits at 90{\%} c.l. on the square of the mixing strength $\left|
{U_{\tau 4} } \right|^2$ between $\nu _\tau $ and a mostly isosinglet heavy neutrino
with mass in the range 10--290 MeV/$c^{2}$  
are reported.  The results were derived using the negative result of a
search for neutral particles decaying into two electrons conducted by the
CHARM collaboration in a neutrino beam dump experiment. Upper limits $ \cong 10^{ - 4}$ were obtained
for neutrino masses larger than 160 MeV/$c^{2}$.}
\newline
\newline
\textit {Keywords :} Neutrino mixing, Neutrino decay
\newpage

Neutrinos may have Dirac or Majorana masses. In general the mass eigenstates 
$\left( {\nu _1, \nu _2, \nu _3 , \nu _4 , \,... } \right)$
 do not 
coincide with the weak (flavour) $( {\nu_e ,\nu_\mu ,\nu_\tau 
,\nu _s ,\,...} )$ eigenstates, but rather with a linear 
combination of them

\begin{equation}
\label{eq:1}
\nu _l = \sum\limits_i {U_{li} \nu _i } \quad \left( {l = e,\,\mu ,\,\tau 
,\,s,\,...;\,i = 1,2,3,4,\,...} \right) \quad\ .
\end{equation}
Such a mixing could result in neutrino oscillations when the mass 
differences are small, and in neutrino decays when the mass differences are 
large. 

In this paper we report limits on the square of the mixing strength $\left| {U_{\tau
4} } \right|^2$ between $\nu _\tau $ and a heavy neutrino, $\nu _4 $, mostly
isosinglet under the Standard $SU(2)_{L}$ gauge group, with mass in
the range 10--290 MeV/$c^{2}$. The limits were
obtained using the negative result of a search for events produced by the
decay of neutral particles into two electrons performed by the
CHARM Collaboration in a neutrino beam dump experiment [1-4]. The
decays of the neutral particles, produced in the dumping of 400 GeV protons
in a Cu target, were looked for in a volume located at a distance
of $L$ = 480 m from the beam dump.

\begin{figure}[htbp]
\includegraphics[width=\hsize]{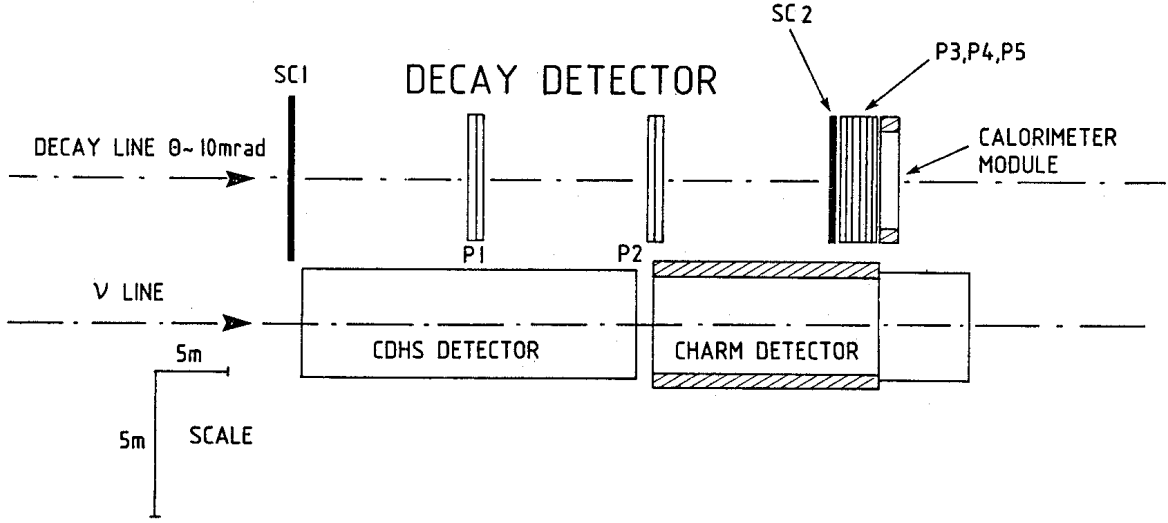}
\caption{Layout of the decay detector.\ \ \ \ \ \ \ \ \ \ \ \ \ \ \ \ \ \ \ \ \
\ \ \ \ \ \ \ \ \ \ \ \ \ \ \ \ \ \ \ \ \ \ \ \ \ \ \ \ \ \ \ \ \ \ \ \ \ \ \ \
\ \ \ \ \ \ \ \ \   }
\label{fig:1}
\end{figure}

The decay detector, shown in Fig.\ref{fig:1}, has already been described elsewhere
[5]. It had an empty decay region of $D$ = 35 m length and 3 m $\times $ 3
m surface area defined by a veto scintillator plane (SC1)
and a scintillator hodoscope (SC2). The volume was subdivided into
three regions using two sets of four proportional tube planes (P1 and
P2) [6]. One module of the CHARM fine-grain calorimeter [6] was displaced
to the end of the decay region. In order to improve the resolution of the
shower angle measurement and to reconstruct better the decay point, three
sets of four proportional tube planes (P3, P4 and P5) were installed
in front of the module. Lead converters of 0.5$X_{0}$ each were placed in
front of P1, P2, P4 and P5. The detector was parallel to the
neutrino beam line at a mean distance of 5 m, corresponding to an angle
with respect to the incident proton beam of 10 mrad, and covered a solid
angle of 3.9 $\times $ 10$^{ - 5}$ sr. The signature of the neutral particles
decaying into two electrons would be events originating in the decay region
at a small angle with respect to the neutrino beam axis with one or two
separate electromagnetic showers.

The detector was exposed to a neutrino flux produced by 1.7 $\times $ 10$^{18 
}$
protons on a solid copper target [7] and 0.7 $\times $ 10$^{18}$ 
protons on a copper target laminated with an effective density of one-third of 
that of solid copper [7]. In the combined exposures, 21 000 events were 
collected satisfying the trigger requirements of no hit in the scintillator 
planes SC1 and a hit in at least four out of the six scintillator planes of the 
calorimeter module. The events were further selected requiring that the 
transverse co-ordinates of the shower vertex lie in a square of 2.5 m $\times 
$ 2.5 m centred on the detector axis and that the electron 
energy, $E_{\rm \tiny el}$, measured in the calorimeter module, be larger than 2 GeV. The 
events recognised as cosmic rays were also rejected. The remaining sample of 7185 
events is dominated by inelastic scattering of electron- and muon-neutrinos 
and antineutrinos producing hadron showers. Compared with the decay of 
neutral particles into two electrons, such events have a broader 
reconstructed angular distribution because of the intrinsic resolution and 
leakage effects. The regularity of the development of electromagnetic 
showers was used to distinguish further between the signal and the 
background events. In particular, the distribution of the deviation of the 
reconstructed direction of the (two) shower(s) from that of the incoming beam and 
the fraction of the energy detected by the proportional drift tubes of the 
calorimeter module outside a narrow cone around the shower were evaluated for the decay 
events by a Monte Carlo method [4]. No event compatible 
with the features of the decay of a neutral particle into two electrons was 
found. 
\newline
\vskip -0.5 cm
\begin{figure}[htbp]
\begin{tabular*}{\hsize}[t]{ccc}
\psfrag{a}[cc][Bc][0.8]{$D_s^+$}
\psfrag{b}[cc][Bc][0.8]{$c$}
\psfrag{c}[cc][Bc][0.8]{$\bar s$}
\psfrag{d}[cc][Bc][0.8]{$U_{\tau4}$}
\psfrag{e}[cc][Bc][0.8]{$\nu_4$}
\psfrag{f}[Bc][Bc][0.8]{$\tau^+$}
\psfrag{g}[Bc][Bc][0.8]{$W^+$}
\includegraphics[width=0.3\hsize]{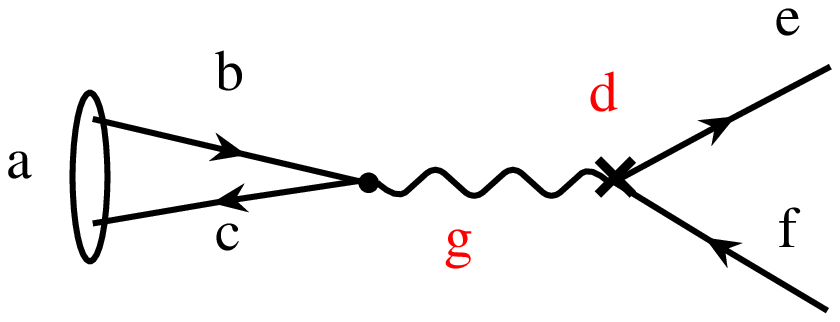} &
\psfrag{a}[c][][0.8]{$\tau^-$}
\psfrag{b}[c][][0.8]{$\nu_4$}
\psfrag{d}[c][][0.8]{$\bar\nu_l$}
\psfrag{c}[c][][0.8]{$l^-$}
\psfrag{e}[c][][0.8]{$U_{\tau4}$}
\psfrag{f}[c][][0.8]{$W^-$}
\includegraphics[width=0.3\hsize]{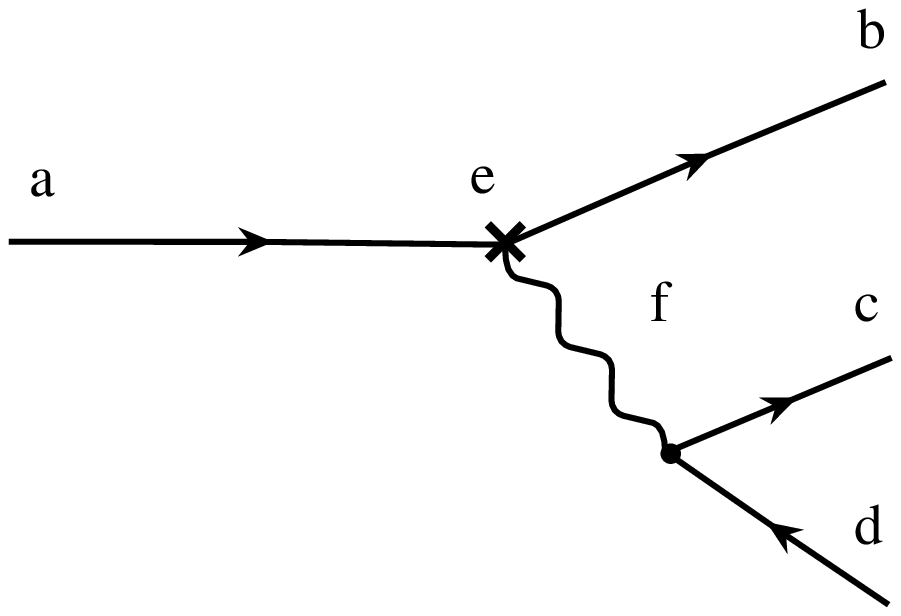}  &
\psfrag{a}[c][][0.8]{$\nu_4$}
\psfrag{b}[c][][0.8]{$\nu_\tau$}
\psfrag{d}[c][][0.8]{$e^+$}
\psfrag{c}[c][][0.8]{$e^-$}
\psfrag{e}[Br][Bc][0.8]{$U_{\tau4}\times$}
\psfrag*{e}[Bl][Bc][0.8][23]{$\sqrt{1-U_{\tau4}^2}$}
\psfrag{f}[c][][0.8]{$Z^0$}
\includegraphics[width=0.3\hsize]{taudecay_1.ps} \\
a) & b) & c) \\ 
\end{tabular*}
\begin{center}
\caption{
Feynman diagrams illustrating: (a) $\nu _4 $ production from
$D_s^+$; (b) leptonic decay of $\tau ^ - $ into a heavy neutrino; (c) decay of an isosinglet
neutrino $\nu _4 $ according to mode (\ref{eq:4}).}
\end{center}
\label{fig:2}
\end{figure}

In this analysis we assumed that the $\nu _4 $ couples to matter essentially 
via mixing with $\nu _\tau $ ($\left| {U_{e4} } \right|^2,\,\left| {U_{\mu 
4} } \right|^2 < < \left| {U_{\tau 4} } \right|^2)$ and we studied the 
proton production of $D_s$'s in the target and their decay into 
$\tau $. The heavy neutrino is produced in the charged current decays

\begin{equation}
\label{eq:2}
D_s \to \nu _4 + \tau 
\end{equation}

and
\begin{equation}
\label{eq:3}
\tau \to \nu _4 + .... \quad\ .
\end{equation}
The Feynman diagrams of the process (\ref{eq:2}) and of the decay $\tau ^ -
\to \nu _4 + l^ - + \bar {\nu }_l $ $ (l = e,\,\mu )$ are shown in
Fig.s 2a and 2b respectively.
On the basis of the assumptions made, the isosinglet heavy neutrino decays 
only via neutral current interactions according to the modes 
\begin{equation}
\label{eq:4}
\nu _4 \to \nu _\tau + {\rm e}^ + + \textrm{e}^ - 
\end{equation}
\begin{equation}
\label{eq:5}
\nu _4 \to \nu _\tau + \nu _l + \bar {\nu }_l \quad \left( {l = \textrm{e},\,\mu 
\,\mbox{and}\,\tau } \right)
\end{equation}
\begin{equation}
\label{eq:6}
\nu _4 \to \nu _\tau + \mu ^ + + \mu ^ - 
\end{equation}
\begin{equation}
\label{eq:7}
\nu _4 \to \nu _\tau + \pi ^0 \quad\ .
\end{equation}
The Feynman diagram illustrating the signal decay channel (\ref{eq:4}) is
shown in Fig. 2c.
The channels (\ref{eq:5})--(\ref{eq:7}) contribute to the
beam attenuation. The branching ratio of mode (\ref{eq:6}) is negligible and
the decay (\ref{eq:7}) opens for neutrino masses larger than the $\pi ^0$
mass. The total decay width is then given by
\begin{eqnarray}
\label{eq:8}
\Gamma _{\rm \tiny tot} = & \Gamma \left( {\nu _4 \to \nu _\tau + \nu _l + \bar {\nu }_l 
} \right) + \Gamma \left( {\nu _4 \to \nu _\tau + e^ + + e^ - } \right) + \\ 
& +\theta \left( {m_{\nu _4 } - m_{\pi ^0} } \right)\,\Gamma \left( {\nu _4 \to 
\nu _\tau + \pi ^0} \right) \nonumber\ .
\end{eqnarray}
%
For heavy neutrinos with mass larger than 290 MeV/$c^{2}$ other decay modes 
open. The leptonic partial width is predicted to be [8]:
\begin{eqnarray}
\label{eq:9}
\Gamma \left( {\nu _4 \to \nu _\tau + \nu _l + \bar {\nu }_l } \right) + 
\Gamma \left( {\nu _4 \to \nu _\tau + e^ + + e^ - } \right) = \\
= K\left[ 
{\frac{\left( {1 + \tilde {g}_{\rm \tiny L}^2 + g_{\rm \tiny R}^2 }
\right)\,G_{\rm \tiny F}^2 m_{\nu _4 }^5 
\left| {U_{\tau 4} } \right|^2\left( {1 - \left| {U_{\tau 4} } \right|^2} 
\right)}{192\pi ^3}} \right] \nonumber\ ,
\end{eqnarray}
%
where $\tilde {g}_{\rm \tiny L} = g_{\rm \tiny L} - 1 = - 1 / 2 + \sin ^2\theta
_{\textrm{w}} $ and $g_{\rm \tiny R} = 
\sin ^2\theta _{\textrm{w}} $, $\theta _{\textrm{w}} $ is the weak angle. In this study the 
neutrinos were assumed to have Dirac masses and then $K$ = 1. For Majorana 
neutrinos $K$ is equal to 2. The leptonic partial width is dominated by the 
mode (\ref{eq:5}):
\begin{equation}
\label{eq:10}
\frac{\Gamma \left( {\nu _4 \to \nu _\tau + \textrm{e}^ + + \textrm{e}^ - } \right)}{\Gamma 
\left( {\nu _4 \to \nu _\tau + \nu _l + \bar {\nu }_l } \right) + \Gamma 
\left( {\nu _4 \to \nu _\tau + \textrm{e}^ + + \textrm{e}^ - } \right)} \cong
0.14\ \ .
\end{equation}
The partial width for the decay (\ref{eq:7}) is predicted to be [9]
\begin{equation}
\label{eq:11}
\Gamma \left( {\nu _4 \to \nu _\tau + \pi ^0} \right) = K\left[
{\frac{G_{\rm \tiny F}^2 
m_{\nu _4 } \left( {m_{\nu _4 }^2 - m_{\pi ^0}^2 } \right)f_\pi ^2 \left| 
{U_{\tau 4} } \right|^2\left( {1 - \left| {U_{\tau 4} } \right|^2} 
\right)}{16\pi }} \right] \quad\ \ . 
\end{equation}

For a given heavy neutrino mass $m_{\nu _4 } $ the number of the decay events 
(\ref{eq:4}) expected in the detector is
\begin{equation}
\label{eq:12}
N = \varepsilon \left( {m_{\nu _4 } } \right)\,\int {\Phi \left( {E_{\nu _4 
} } \right)} \,P_{\nu _4 \to \nu _\tau e^ + e^ - } \left( {E_{\nu _4 } } 
\right)\,\,dE_{\nu _4 } \quad\ ,
\end{equation}
where $\Phi \left( {E_{\nu _4 } } \right)$ is the differential
 flux of heavy neutrinos,
 $P_{\nu _4 \to \nu _\tau e^ + e^ - } \left( {E_{\nu _4 } } 
\right)$ is the probability for a heavy neutrino of energy $E_{\nu _4 } $ to 
decay in the decay fiducial volume according to reaction (\ref{eq:4}), and 
$\varepsilon \left( {m_{\nu _4 } } \right)$ is the efficiency of the 
selection criteria of one or two electrons in the calorimeter based on the 
regularity of the development of electromagnetic showers and the 
collinearity between the (two) shower(s) and neutrino direction. The flux 
$\Phi \left( {E_{\nu _4 } } \right)$ is given by
\begin{equation}
\label{eq:13}
\Phi \left( {E_{\nu _4 } } \right) = N_{\rm \tiny p} \frac{\sigma _{D_s } }{\sigma 
_{\rm \tiny inel} }\left[ {\begin{array}{l}
 \textrm{BR}\left( {D_s \to \nu _4 + \tau } \right)A_{\nu _4 }^{D_s } \,\phi _{\nu _4 
}^{D_s } \left( {E_{\nu _4 } } \right) + \\ 
 + \textrm{BR}\left( {D_s \to \nu _\tau + \tau } \right)\textrm{BR}\left( {\tau \to \nu _4 + 
...} \right)A_{\nu _4 }^\tau \phi _{\nu _4 }^\tau \left( {E_{\nu _4 } }
\right)
 \end{array}} \right]\ \ .
\end{equation}

The number of protons on the target corrected for the detector dead time, 
13.6{\%} for the solid target and 21.7{\%} for the laminated one, is 
$N_{\rm \tiny p}$ = 2.0$\times $10$^{18}$. The fraction of proton inelastic 
interactions leading to a charged $D_{s}$ is given by [10]
\begin{equation}
\label{eq:14}
\frac{\sigma _{D_s } }{\sigma _{\rm \tiny inel} } = \frac{A_{\rm \tiny Cu} \left[ {{\sigma 
\left( {D_s } \right)} \mathord{\left/ {\vphantom {{\sigma \left( {D_s } 
\right)} {\sigma \left( D \right)}}} \right. \kern-\nulldelimiterspace} 
{\sigma \left( D \right)}} \right]\sigma _D^{\rm \tiny nucleon} }{\sigma
_{\rm \tiny inel}^{\rm \tiny Cu} 
} = 2.98\times 10^{ - 4}\ \ ,
\end{equation}
where the copper mass number is $A_{\rm \tiny Cu} = 63.55$. Linear $A$ dependence is
assumed for charm production. The used values of the inelastic proton cross--section, 
$\sigma _{\rm \tiny inel}^{\rm \tiny Cu} $ [11], of the ratio of the production
cross--section for $D_s^\pm $ over the production cross--section for $D^\pm +
D^0$, ${\sigma \left( {D_s } \right)} \mathord{\left/ {\vphantom {{\sigma
\left( {D_s } \right)} {\sigma \left( D \right)}}} \right.
\kern-\nulldelimiterspace} {\sigma \left( D \right)}$ [12], and of the
inclusive cross--sections for the production of $D$ mesons, $\sigma
_D^{\rm \tiny nucleon} $ [13], are reported in Table 1. The ratio ${\sigma \left( {D_s
} \right)} \mathord{\left/ {\vphantom {{\sigma \left( {D_s } \right)}
{\sigma \left( D \right)}}} \right. \kern-\nulldelimiterspace} {\sigma
\left( D \right)}$ was obtained by the Beatrice experiment studying charmed
particles produced by $\pi^{ - }$'s of 350 GeV/$c$ [12]. It is
compatible with the results obtained by $\textrm{e}^{ + }\textrm{e}^{ - }$ experiments at
center of mass energies equal to 10 GeV and at $\textrm{Z}^{0}$ mass [14]. The
value of $\sigma _D^{\rm \tiny nucleon} $ was obtained by the NA27 Collaboration
studying the production of $D$'s in the interactions of 400 GeV protons in
an $\textrm{H}_{2}$ target [13].
\newline
\newline

\begin{table}[htbp]
\begin{center}
\begin{tabular}{|l|c|c|}
\hline
Parameters& 
Values& 
Systematic\goodbreak errors [{\%}] \\
\hline
$\sigma _{\rm \tiny inel}^{\rm \tiny Cu} \left[ {\mbox{mb}} \right] $ [11]& 
$769\pm 23$\ \ & 
\ 3.0 \\
\hline
${\sigma \left( {D_s } \right)} \mathord{\left/ {\vphantom 
{{\sigma \left( {D_s } \right)} 
{\sigma \left( D \right)}}} \right. 
\kern-\nulldelimiterspace} {\sigma \left( D \right)}$ [12]& 
$0.12\pm 0.03$& 
25.0 \\
\hline
$\sigma _D^{\rm \tiny nucleon} \left[ {\mu {\kern 1pt} 
{\kern 1pt} \mbox{b}} \right]$ [13]& 
$30.1\pm 3.1$\ \ & 
10.3  \\
\hline
$Br\left( {D_s \to \nu _\tau + \tau } \right)$ [15]& 
$0.07\pm 0.04$& 
57.1  \\
\hline
$n$ [13]& 
$4.9\pm 0.5$& 
\ 4.0 \\
\hline
$b$ [13]& 
$1.0\pm 0.1$& 
\ 5.0 \\
\hline
Spectra of $\nu _4 $ \par produced in $\tau $ decay& 
see text& 
\ 5.0 \\
\hline
\end{tabular}
\caption{Values of the parameters used in the analysis and their
contribution in percentage to the systematic error on the expected number
of decay events (\ref{eq:4}). The uncertainty due to the knowledge of the
spectra of $\nu _4 $ produced in $\tau $ decay refers to heavy neutrinos
from reaction (\ref{eq:3}).}
\label{tab1}
\end{center}
\end{table}
\begin{table}[htbp]
\begin{center}
\begin{tabular}{|c|l|c|}
\hline
$i$& 
Mode& 
Branching ratio $\left[ \% \right]$ \\
\hline
1.& 
$\tau \to \mu + \nu _\mu + \nu _\tau $& 
17.37\ \  \\
\hline
2.& 
$\tau \to e + \nu _e + \nu _\tau $& 
17.83\ \  \\
\hline
3.& 
$\tau \to \pi ^ - + \nu _\tau $& 
11.09\ \  \\
\hline
4.& 
$\tau \to \pi ^ - + \pi ^0 + \nu _\tau $& 
25.40\ \  \\
\hline
5.& 
$\tau \to \pi ^ - + K^0 + \nu _\tau $& 
1.06 \\
\hline
6.& 
$\tau \to \pi ^ - + 2\pi ^0 + \nu _\tau $& 
9.13 \\
\hline
7.& 
$\tau \to \pi ^ + + 2\pi ^ - + \nu _\tau $& 
9.49 \\
\hline
8.& 
$\tau \to \pi ^ - + 3\pi ^0 + \nu _\tau $& 
1.21 \\
\hline
9.& 
$\tau \to 2\pi ^ - + \pi ^ + + \pi ^0 + \nu _\tau $& 
4.32 \\
\hline
10.& 
$\tau \to 2\pi ^ - + \pi ^ + + K^0 + \nu _\tau $& 
1.35 \\
\hline
11.& 
$\tau \to e + \gamma + \nu _e + \nu _\tau $& 
1.75 \\
\hline
\end{tabular}
\caption{The $\tau $ decay modes and corresponding branching ratio values used in 
the analysis \ \ \ \ \ \ \ \ \ \ \ \ \ \ \ \ \ \ \ \ \ \ \ \ \ \ \ \ }
\label{tab2}
\end{center}
\end{table}
In equation
(\ref{eq:13}) one has
\begin{equation}
\label{eq:15}
\textrm{BR}\left( {D_s \to \nu _4 + \tau } \right) = \textrm{BR}\left( {D_s \to \nu _\tau + 
\tau } \right)\rho _{D_s } \left| {U_{\tau 4} } \right|^2\ \ .
\end{equation}
The value of the branching ratio of
the $D_{s}$ decay into a zero mass neutrino, \linebreak $\textrm{BR}\left( {D_s \to
\nu _\tau + \tau } \right)$, is reported in Table 1 [15]. Its uncertainty
dominates the systematic error of this study. The factor $\rho_{D_s}$
describes
phase space and helicity effects [16]. In the
case of $\tau $ decay into $\nu _4 $ 
\begin{equation}
\label{eq:16}
\textrm{BR}\left( {\tau \to \nu _4 + ...} \right) = \left| {U_{\tau 4} } 
\right|^2\sum\limits_i {\textrm{BR}\left( {\tau \to \nu _\tau + X_i } \right)\rho 
_\tau ^i = } \rho _\tau \left| {U_{\tau 4} } \right|^2\ \ ,
\end{equation}
where $\textrm{BR}\left( {\tau \to \nu _\tau + X_i } \right)$ is the branching ratio 
of the $\tau $ decay into a zero mass neutrino according to the considered 
mode $i$ [15], see Table 2, and the $\rho _\tau ^i $ are factors depending on 
heavy neutrino mass. For $i$ = 1--3 they take into account
phase space and 
helicity effects [17]. For the modes 4--11 the $\rho _\tau ^i $'s 
were computed using only phase space [18]. The numerical values of $\rho 
_\tau ^4 $ as a function of the neutrino mass are smaller than the ones of 
Ref. [19] by about 10{\%}. The latter were obtained taking into account also helicity effects
and the experimental width of the vector meson $\rho$. 

For a given neutrino mass, the acceptances of the heavy neutrino flux
coming from the $D_{s}$ [\textit{$\tau $}], $A_{_{\nu _4 } }^{D_s } $
[$A_{_{\nu _4 } }^\tau $], and the corresponding energy spectrum normalized
to one, $\phi _{\nu _4 }^{D_s } \left( {E_{\nu _4 } }\right) $
[$\phi _{\nu _4 }^\tau \left( {E_{\nu _4 } }\right) $]
were obtained using a Monte Carlo simulation. The production of strange charm by
protons was parametrized using the
semi-empirical expression
\begin{equation}
\label{eq:17}
f\left( {x_F } \right) \approx \left( {1 - \left| {x_F } \right|} 
\right)^ne^{ - bp_{\mbox{\rm \tiny T}^2 }}\ \ ,
\end{equation}
where $x_{F}$ is the meson longitudinal momentum in the collision
center of mass frame divided by its maximum value ${\sqrt s }
\mathord{\left/ {\vphantom {{\sqrt s } 2}} \right.
\kern-\nulldelimiterspace} 2$, and $p_{\rm \tiny T}$ is the meson transverse momentum.
Since there are few experimental results available on the production of
$D_s^\pm $, the values of $n$ and $b$ were inferred from the measurements
of $D$ production. Assuming the hadronization process to be independent of the
$\textrm{c}\bar {\textrm{c}}$ production mechanism, the parameters $n$ and $b$ are
independent of the meson produced. Most measurements agree with a value of
$b$ equal to 1 (GeV/$c$)$^{ - 2}$. The used values reported in Table 1 were
obtained by the NA27 Collaboration studying the production of $D$'s in the
interactions of 400 GeV protons in an $\textrm{H}_{2}$ target [13]. Cascade
production was neglected.

The energy spectrum of heavy neutrinos from \textit{$\tau $ }decay is given by 
\begin{equation}
\label{eq:18}
\phi _{\nu _4 }^\tau \left( {E_{\nu _4 } } \right) = \frac{\sum\limits_{i = 
1}^{11} {\rho _\tau ^i \phi _{\nu _4 }^i \left( {E_{\nu _4 } } \right)} 
}{\sum\limits_{i = 1}^{11} {\rho _\tau ^i } }\ \ ,
\end{equation}
where $\phi _{\nu _4 }^i \left( {E_{\nu _4 } } \right)$ is the normalized
energy distribution of neutrinos produced in the decay mode $i$ (see Table
2).  In the case of leptonic channels, $i$ = 1 and 2, the spectra were
obtained using the matrix element
\begin{equation}
\label{eq:19}
\left| A \right|^2 \approx \left( {p_\tau \cdot p_{\nu _l } } 
\right)\,\left( {p_l \cdot p_{\nu _4 } } \right)\ .
\end{equation}
The quantities $p_\tau ,\,p_{\nu _l } ,\,p_l$ and $p_{\nu _4 }$ are
the four-momenta of $\tau $, light neutrino, electron or muon, and
heavy neutrino, respectively. The spectrum of heavy neutrino produced by
channel 11 was obtained using phase space. The multi--pion decay modes were
simulated using two models. In model (a) the spectra of channels 4--10 and
their relative contributions as a function of the heavy neutrino mass were
computed using phase space [18]. In model (b) channel 4 was assumed to be
produced through the resonance \textit{$\rho $} and channels 5--10 through the
resonance $a_{1}$. The resonances were assumed to have zero width. The $\rho
_\tau ^i $'s in (\ref{eq:18}) are given by:
\begin{equation}
\label{eq:20}
\rho _\tau ^i = \frac{\left( {1 - y} \right)^2 + x\left( {1 + y - 2x} 
\right)}{1 + x - 2x^2}\sqrt {1 - y\left[ {\frac{2 + 2x - y}{\left( {1 - x} 
\right)^2}} \right]}\ \ ,
\end{equation}
where $x = {m_\rho ^2 } \mathord{\left/ {\vphantom {{m_\rho ^2 } {m_\tau ^2 
}}} \right. \kern-\nulldelimiterspace} {m_\tau ^2 }$ for $i$ = 4, ($m_{\rho 
}$ = 770 MeV/$c^{2})$, $x = {m_{a_1 }^2 } \mathord{\left/ {\vphantom {{m_{a_1 
}^2 } {m_\tau ^2 }}} \right. \kern-\nulldelimiterspace} {m_\tau ^2 }$ for 
$i$ = 5--10 ($m_{a_1 } $ = 1260 MeV/$c^{2})$ and $y = {m_{\nu _4 }^2 } 
\mathord{\left/ {\vphantom {{m_{\nu _4 }^2 } {m_\tau ^2 }}} \right. 
\kern-\nulldelimiterspace} {m_\tau ^2 }$. The values of Table 3 were 
computed using the average of the spectra obtained in the two models. The 
systematic error in Table 1 reflects the differences of the spectra.

The decay (\ref{eq:4}) was simulated using the matrix element [8]
\begin{equation}
\label{eq:21}
\left| A \right|^2 \approx \left[ {\tilde {g}_L^2 \left( {p_{\nu _4 } \cdot 
p_{e^ - } } \right)\left( {p_{\nu _\tau } \cdot p_{e^ + } } \right) + g_R^2 
\left( {p_{\nu _4 } \cdot p_{e^ + } } \right)\left( {p_{\nu _\tau } \cdot 
p_{e^ - } } \right)} \right]
\end{equation}
which neglects the electron mass. The quantities $p_{e^ - } ,\,p_{e^ + }$
and $p_{\nu _\tau }$ are the four-momenta of electron, positron and tau
neutrino, respectively. In the center of mass of the decaying heavy neutrino
the four-vector $p_{\nu _4 }$ is given by [20]
\begin{equation}
\label{eq:22}
p_{\nu _4 } = \left( {m_{\nu _4 } ,\, - m_{\nu _4 } \left| h \right|\vec 
{\eta }} \right)\ \ ,
\end{equation}
where $\vec {\eta }$ is a unit vector parallel to the direction of the
heavy neutrino in the rest frame of the particle decaying into neutrino,
$D_s $, or $\tau $, and $\left| h \right|$ is the absolute value of the
neutrino (antineutrino) helicity. In the case of heavy neutrinos coming
from $D_{s}$ decay, the values of $\left| h \right|$ obtained in Ref. [15]
were used. As the polarization of $\tau $ produced in $D_s $ decay
is negligible, $\left| h \right| = 0$ was assumed for the heavy neutrinos
produced in $\tau $ decay. The acceptances and the mean momenta of
decaying heavy neutrinos expected to be detected in the detector are
reported in Table 3 for different values of neutrino mass. The efficiency
of the cut $E_{\rm \tiny el} > 2$ GeV is about 85{\%} for heavy neutrinos coming from
$D_s $ and larger than 95{\%} for the ones coming from \textit{$\tau $}.

In Eq. (\ref{eq:12}) the probability for a heavy neutrino of energy
$E_{\nu _4 } $ to decay in the decay fiducial volume is given by
\begin{equation}
\label{eq:23}
P_{\nu _4 \to \nu _\tau {\rm \tiny e}^ + {\rm \tiny e}^ - } \left( {E_{\nu _4 } } \right) = e^{ - 
\,\,\frac{L}{\lambda }}\left( {1 - e^{ - \,\,\frac{D}{\lambda }}} 
\right)\frac{\Gamma \left( {\nu _4 \to \nu _\tau + {\rm e}^ + + {\rm e}^ - } 
\right)}{\Gamma _{\rm \tiny tot} }\ ,
\end{equation}
where $\lambda = {\left( {\gamma \beta {\kern 1pt} c} \right)}
\mathord{\left/ {\vphantom {{\left( {\gamma \beta {\kern 1pt} c} \right)}
{\Gamma _{tot} }}} \right. \kern-\nulldelimiterspace} {\Gamma _{\rm \tiny tot} }$ is
the heavy neutrino mean decay path ($\gamma = {E_{\nu _4 } }
\mathord{\left/ {\vphantom {{E_{\nu _4 } } {m_{\nu _4 } }}} \right.
\kern-\nulldelimiterspace} {m_{\nu _4 } }$, $\beta = {p_{\nu _4 } }
\mathord{\left/ {\vphantom {{p_{\nu _4 } } {E_{\nu _4 } }}} \right.
\kern-\nulldelimiterspace} {E_{\nu _4 } }$).  According to Eqs.
(\ref{eq:9}) and (\ref{eq:11}) $\lambda $ depends on $\left| {U_{\tau 4} }
\right|^2\left( {1 - \left| {U_{\tau 4} } \right|^2} \right)$.

The quantity $\varepsilon \left( {m_{\nu _4 } } \right)$ in Eq.
(\ref{eq:12}) is the efficiency of the selection criteria based on the
regularity of the development of electromagnetic showers and the
collinearity between the (two) shower(s) and the neutrino direction. The
values, obtained using a detailed Monte Carlo simulation of the detector
response, decrease with increasing heavy neutrino mass. It ranges from
91{\%} for $m_{\nu _4 } = 10 \,\mbox{MeV} /c^2$ to 65{\%} for $m_{\nu _4 } =
290 \,\mbox{MeV} / c^2$ [4].

\begin{table}[htbp]
\begin{center}
\begin{tabular} {|c|c|c|c|c|}
\hline
$m_{\nu _4 } \left[ {\,\mbox{MeV} / c^2} \right]$& 
$A_{\nu _4 }^{D_s } \times 10^{ - 3}$& 
$\left\langle {p_{\nu _4 }^{D_s } } \right\rangle \,\left[ {\,\mbox{GeV}} \right]$& 
$A_{\nu _4 }^\tau \times 10^{ - 3}$& 
$\left\langle {p_{\nu _4 }^\tau } \right\rangle \,\left[ {\,\mbox{GeV}} \right]$ \\
\hline
\ \ 10& 
3.39& 
14.39& 
3.49& 
47.14 \\
\hline
\ \ 50& 
3.50& 
14.14& 
3.54& 
47.13 \\
\hline
100& 
4.11& 
12.90& 
3.61& 
46.86 \\
\hline
150& 
6.00& 
10.39& 
3.73& 
46.48 \\
\hline
190& 
8.00& 
\ \ 8.38& 
3.87& 
46.44 \\
\hline
250& 
-& 
\ \ -& 
4.04& 
46.03 \\
\hline
290& 
-& 
\ \ -& 
4.24& 
45.62 \\
\hline
\end{tabular}
\caption{Acceptances and mean momenta of decaying heavy neutrinos 
with $E_{\rm \tiny el} > 2$ GeV for different values of the neutrino 
mass.}
\label{tab3}
\end{center}
\end{table}

Since no decay event was detected, upper limits at 90{\%} confidence level 
on $\left| {U_{\tau 4} } \right|^2$ were obtained in the neutrino mass range 
10--290 MeV/$c^{2}$. The limit value of $N = N_l = 6.42$ events was used in 
Eq. (\ref{eq:12}). Since the contribution to the systematic error of the 
uncertainty on the spectra of heavy neutrinos coming from $\tau $ decay is 
negligible, $N_l$ does not depend on the neutrino mass. It corresponds to an 
average probability of observing no events, $\left\langle {P_0 \left( {N_l } 
\right)} \right\rangle $, equal to 10{\%} [21]:
\begin{equation}
\label{eq:24}
\left\langle {P_0 \left( {N_l } \right)} \right\rangle =
 \int\limits_{ -\infty}^{ +\infty } {f\left( {0;\,N_l '} \right)\,W\left( {N_l ';N_l ,\sigma } 
\right)\,dN_l '} = 0.1\ ,
\end{equation}
where $f\left( {0;\,N_l '} \right) = e^{ - N_l '}$ is the Poisson
probability of obtaining zero events. In the case of negative $N_{l}'$, 
$f\left( {0;\,N_l '} \right) = 1$ was used. The probability density function
$W\left( {N_l ';N_l ,\sigma } \right)$ takes into accounts the systematic
errors summarized in Table 1 and was assumed to be a Gaussian
distribution. Combining in quadrature the uncertainties reported in the table, one
gets $\sigma \mathord{\left/ {\vphantom {\sigma {N_l }}} \right.
\kern-\nulldelimiterspace} {N_l } = 0.64$. In the case of no uncertainty, $W\left( {N_l ';N_l
,\sigma } \right) = \delta \left( {N_l ' - N_l } \right)$ and the integral
gives $N_l = 2.30$: the upper limit at 90{\%} confidence level for the mean
value of a Poisson distribution in the case of zero observations. The
chosen value $N_l = 6.42$ is a safely conservative number: taking for
instance $W\left( {N_l ';N_l
,\sigma } \right)$ equal to a
log-normal distribution to avoid negative values of $N_l'$ [21], 
one obtains $N_l=3.21$, a factor 2 smaller than the value used in the analysis.

\begin{figure}[htbp]
\psfrag{1E0}[Bc][Bc]{$10^{0}$\ \ \ \ }
\psfrag{1E-1}[Bc][Bc]{$10^{-1}$}
\psfrag{1E-2}[Bc][Bc]{$10^{-2}$}
\psfrag{1E-3}[Bc][Bc]{$10^{-3}$}
\psfrag{1E-4}[Bc][Bc]{$10^{-4}$}
\psfrag{1E-5}[Bc][Bc]{$10^{-5}$}
\psfrag{U2}[Bc][Bc]{$|U_{\tau 4}|^2$}
\psfrag{mnu4(MeV)}[Bc][Bc]{$m_{\nu_4}$ (MeV)}
\centerline{\includegraphics[width=\hsize]{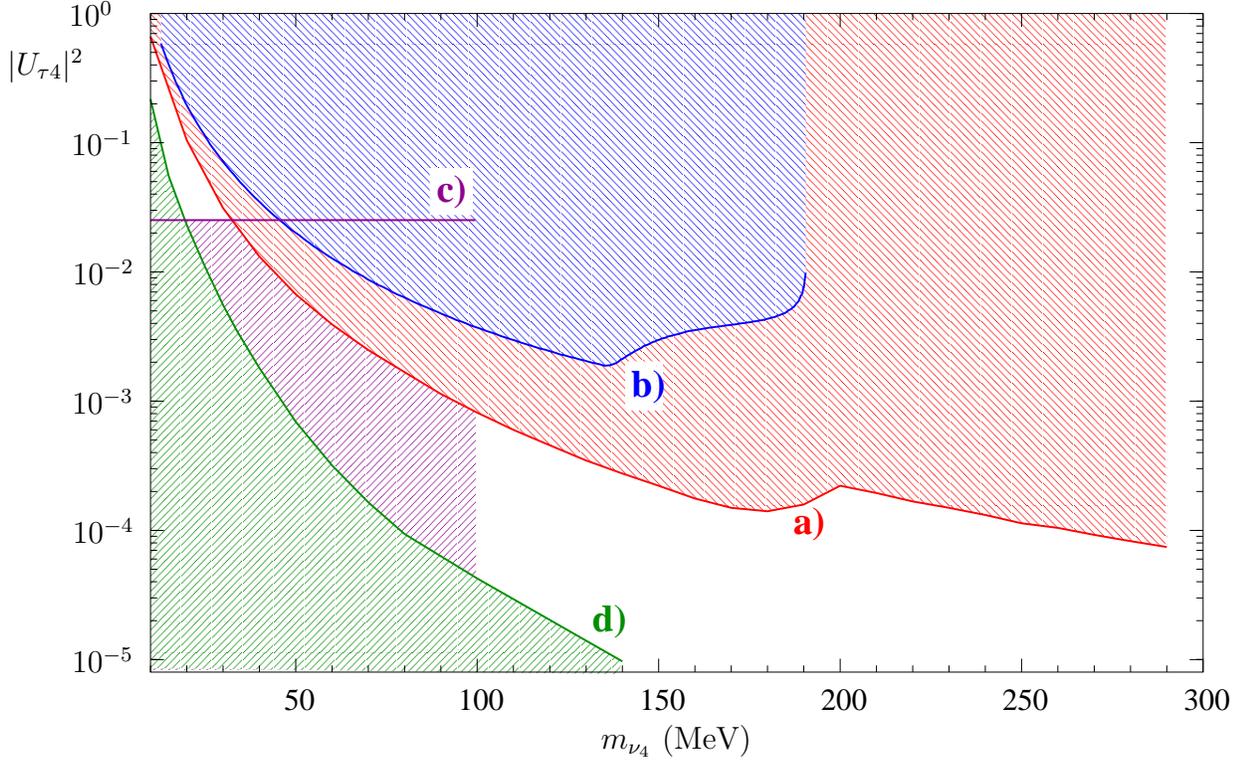}}
\caption{Limits at 90{\%} confidence level on the square of the mixing strength $\left| {U_{\tau 4} }
\right|^2$ of the $\tau$-neutrino
with a fourth neutrino mass eigenstate mostly isosinglet : (a) upper limits from this study;
(b) the NOMAD upper limits [22]. The lower limits (c) from
SN1987a and (d) from the Big Bang Nucleosynthesis constraint
$\Delta N_\nu ^{eff} \le 0.2$ are reproduced from Ref. [9]. Upper limits of
$\left| {U_{\tau 4} } \right|^2 \le 10^{ - 8}$ from SN1987a and
of $\left| {U_{\tau 4} } \right|^2 \le 10^{ - 10} - 10^{ - 12}$ from Big
Bang Nucleosynthesis have also been derived for the corresponding mass
range shown, respectively [9]. All the limits were obtained assuming that
neutrinos have Dirac masses ($K$ = 1).}
\label{fig:3}
\end{figure}

The upper limits obtained at 90{\%} confidence level on $\left| {U_{\tau 4} 
} \right|^2$ values, as a function of $m_{\nu _4 } $, are shown in
Fig.\ref{fig:3}, 
together with previous results [9, 22]. Limits on $\left| {U_{\tau 4} } 
\right|^2$ were also obtained for neutrino masses larger than $140\,\mbox{MeV} / c^2$ 
assuming $\left| {U_{e4} } \right|^2 = \left| {U_{\mu 4} } 
\right|^2 = \left| {U_{\tau 4} } \right|^2$ from the upper bounds on the rates of the 
decays $\tau ^ - \to e^\pm \left( {\mu ^\pm } \right)\,{\kern 1pt} \pi ^ \mp 
\pi ^ - $  [23]. Limits on $\left| {U_{e4} 
} \right|^2$ and $\left| {U_{\mu 4} } \right|^2$ are reported in Ref. 
[15]. 

In conclusion, the negative results of a search of decays of neutral 
particles into two electrons performed by the CHARM Collaboration in a neutrino 
beam dump experiment, allowed limits to be set at 90{\%} c.l. on the square of 
the mixing strength, $\left| {U_{\tau 4} } \right|^2$, between $\nu _\tau $ and a 
mostly isosinglet fourth neutrino, $\nu _4 $, having a mass in 
the range 10--290 MeV/$c^{2}$. Values of $ \cong $10$^{ - 4 }$ were obtained 
for masses larger than $160\,\mbox{MeV} / c^2$.

We thank the members of the CHARM Collaboration who 
allowed us to use their data. We would like to thank B. Van de Vyver for 
the program simulating the generation of prompt neutrinos. We thank G. 
Barbiellini, L. Di Lella, A.D. Dolgov, S.N. Gninenko, P. Loverre, M. Mangano, J.
Swain 
and K. Winter for many interesting discussions and comments. We would also like 
to thank S.H. Hansen for advice on the astrophysical 
and cosmological limits.
 

\subsubsection*{References}

\begin{enumerate}

\item F. Bergsma et al., (CHARM Collaboration), Phys. Lett. 128B (1983) 361.

\item J. Dorenbosch et al., (CHARM Collaboration), Phys. Lett. 166B (1986) 473.

\item F. Bergsma et al., (CHARM Collaboration), Phys. Lett. 157B (1985) 458.

\item J. Aspiazu, Suche nach Zerf\"{a}llen neutraler durchdringender Teilchen, 
Dissertation zur Erlangung des Doktorgrades des Fachbereichs Physik der 
Universit\"{a}t Hamburg, II Institut f\"{u}r Experimentalphysik, 
Universit\"{a}t Hamburg, 1985.

\item M. Jonker et al., (CHARM Collaboration), CERN/SPSC/81-21, SPSC/P142 Add1, 12 
March 1981.

\item M. Jonker et al., (CHARM Collaboration), Nucl. Instrum. Methods 200 
(1982) 183.

\item M. Jonker et al., (CHARM Collaboration), CERN/SPSC/80-31, SPSC/P142 Add1, 25 
April 1980 (Experiment WA65).

\item A.D. Dolgov et al., Nucl. Phys. B 580 (2000) 331.

\item A.D. Dolgov et al., Nucl. Phys. B 590 (2000) 562.

\item B. Van de Vyver, Nucl. Instrum. Methods Phys. Res. A 385 
(1997) 91.

\item A.S. Carroll et al., Phys. Lett. 80B (1979) 319.

\item M. Adamovich et al., (Beatrice Collaboration), Nucl. Phys. B 495 (1997) 
3.

\item M. Aguilar-Benitez et al., (LEBC-EHS Collaboration), Phys. Lett. 189B 
(1987) 476.

\item L. Gladilin, Charm hadron production fraction, hep-ex/9912064 (1999).

\item D.E. Groom et al., Particle Data Group, Eur. Phys. J. C15, 1 
(2000).

\item R.E. Shrock, Phys. Rev. D24 (1981) 1232.

\item J. Swain and L. Taylor, Phys. Rev. D 55 (1997) R1.

\item W.S.C. Williams, An Introduction to elementary particles, Second Edition, 
Academic Press 1971, Appendix C, p. 499.

\item S. Narison, Z. Phys. C 2, 11 (1979).

\item L.B. Okun, Leptons and Quarks, 
North--Holland Publishing Company,
\linebreak 
Amsterdam--New York--Oxford, 1989, p. 327.

\item R.D. Cousins and V.L. Highland, Nucl. Instrum. Methods Phys. 
Res. A320 (1992) 331.

\item P. Astier et al., (NOMAD Collaboration), Phys. Lett. 506B (2001) 27.

\item V. Gribanov, S. Kovalenko and I. Schmidt, Nucl. Phys. B 607 (2001) 335.
\end{enumerate}

\end{document}